# A Nickelate Renaissance

J.F. Mitchell
Materials Science Division
Argonne National Laboratory
Lemont, IL. 60439

Abstract

The 2019 discovery of high temperature superconductivity in layered nickelate films, $Nd_{1-x}SrNiO_2$, has galvanized a community that has been studying nickelates for more than 30 years both as cuprate analogs and in their own right. On the surface, infinite layer nickelates, and their multilayer analogs, should be promising candidates based on our understanding of cuprates: square planar coordination and a parent $d^9$ configuration that places a single hole in a $dx^2$-$y^2$ planar orbital makes nickelates seem poised for superconductivity. But creating crystals and films of sufficient quality of this $d^9$ configuration in $Ni^{1+}$ has proven to be a synthetic challenge, only recently overcome. These crystalline specimens are opening windows that shed new light on the cuprate-nickelate analogy and revealing nuances that leave the relationship between cuprates and nickelates very much an area open to debate further study. This Perspective gives a qualitative, phenomenological account of these newly discovered superconductors and multilayer members of the infinite layer nickelate family. The focus is on our current understanding of electronic and magnetic properties of these materials as well as some future opportunities, explored from the viewpoint of synthetic challenges and some suggested developments in materials discovery and growth to make further progress in this rejuvenated field.

**Introduction**

"Like the soil, mind is fertilized while it lies fallow, until a new burst of bloom ensues[1]. These words were penned by John Dewey in his 1934 work on the philosophy of art, but they express a universal experience that is common across most walks of life, especially the sciences. The materials that inspire the collection of articles in this issue, low-dimensional rare earth nickelates, are a case in point. With a story spanning more than three decades, these compelling materials have been fellow travelers with the more famous copper oxide superconductors, yet have always remained in their shadow, a 'poor cousin.' That is, until now.

With superconductivity discovered in epitaxial thin films of the 'infinite-layer' $d^9$ nickelate (Nd,Sr)NiO$_2$ by Harold Hwang[2] and his Stanford collaborators in mid-2019 and more recently by Julia Mundy[3] of Harvard in the five-layer nickelate Nd$_6$Ni$_5$O$_{12}$, a decades-long dream has been realized. Their discoveries and other exciting new findings in $d^9$ layered nickelate materials rest on breakthroughs in thin film and single crystal growth that have allowed old ideas to take on a new life, and renewed enthusiasm to explore the nickelates as a platform to understand high-$T_c$ superconductivity beyond cuprates, indeed a 'new burst of bloom.' The purpose of this Perspective is to provide a phenomenological account of how this nickelate renaissance has emerged as viewed through the lens of new and transformative materials developments in $d^9$ nickel oxides, the discoveries they have enabled, and some new directions toward which they may point.

*Cuprates*

To appreciate the long-lived fascination with nickelates and the importance of today's resurgence of interest, a bit of stage-setting about cuprate superconductors is essential. Understanding the physics behind the copper oxides has been a defining challenge of condensed matter physics for the 35 years since their discovery. The hundreds of thousands of papers written about cuprates underscore simultaneously their impact on the condensed matter community, and the daunting challenge to untangling both the phenomenology and mechanism behind these unconventional superconductors. Generating this challenge is an elaborate web of interactions that result in the complex phase behavior of cuprates, behavior that contains—in addition to superconductivity—magnetism, charge order, non-Fermi-liquid metals, real-space stripes (both static and dynamic) pair density waves, and perhaps other yet-to-be-discovered exotica.[4] Unraveling how these emergent phases compete or cooperate, both in the normal and superconducting state, has driven the field from its inception to today.

Superconducting cuprates emerge by doping a parent antiferromagnetic insulating parent oxide containing $Cu^{2+}$, exemplified by La$_2$CuO$_4$ (La-214). Doping introduces holes (i.e., La$_{2-x}$Sr$_x$CuO$_4$) or electrons (i.e., Nd$_{2-x}$Ce$_x$CuO$_4$) to produce first a metal, and then a superconductor. Along with sophisticated quantitative models and mechanisms constructed to explain this progression of phases, there have emerged a set of phenomenological descriptors that are considered important 'ingredients' in the 'recipe' for high-$T_c$ copper oxide superconductivity. These include:

- A quasi-2D structure in which the copper ions are surrounded by a highly axially-elongated octahedral coordination polyhedron of oxygen anions.
- An orbital configuration around Cu in which the $dx^2$-$y^2$ orbital is highest lying. With a nominal oxidation state of Cu$^{2+}$, this $d^9$ filling promotes a Jahn-Teller distortion that generates a configuration with a single hole in this half-filled $dx^2$-$y^2$ orbital.
- Strong correlations precipitate an instability in the notionally metallic half-filled $dx^2$-$y^2$ band of the parent, leading to an insulating, magnetic ground state (so-called 'checkerboard' antiferromagnetic order)

- The relatively narrow separation between Cu $3d$ and O $2p$ states implies a considerable hybridization between these orbitals and leads to a significant O component to the hole wave function, placing cuprates solidly in the charge transfer insulator rather than the Mott regime of the Zaanen-Sawatzky-Allen framework.[5]

## A *Cuprate Analog*

These ingredients have driven materials and chemistry design rules for high-$T_c$ superconductors in the search both for other cuprate families and for non-copper based analogs. Soon after the discovery of cuprate superconductivity, attention turned to known nickelates isostructural with La-214, specifically $La_{2-x}Sr_xNiO_4$ (LSNO). Although Ni is immediately to the left of Cu on the Periodic Table, and a similar solid-state chemistry is found, the parent phase contains $Ni^{2+}$, $d^8$, rather than $d^9$. Such $d^8$ systems in octahedral coordination are typically high spin with holes in both the $dx^2$-$y^2$ and $dz^2$ orbitals. Additionally, the lower electronegativity of Ni vs Cu means that the separation between the O 2p states and the Ni $d$ manifold is larger, reducing hybridization. With these cuprate ingredients missing, it is perhaps unsurprising that superconductivity has not been found in LSNO. Indeed, the system remains insulating unless doped heavily to x≈ 0.9,[6] where a metal emerges. Study of this system and analogs has revealed collective phenomena interesting on their own right, such as real-space charge and spin stripes,[7-12] whose period is set by the charge concentration, e.g., $La_{5/3}Sr_{1/3}NiO_4$ with 1/3 hole added to $Ni^{2+}$ has a three-fold superlattice oriented at 45° to the Ni-O bonds. Many of these phenomena turn out to be relevant to cuprates as well.

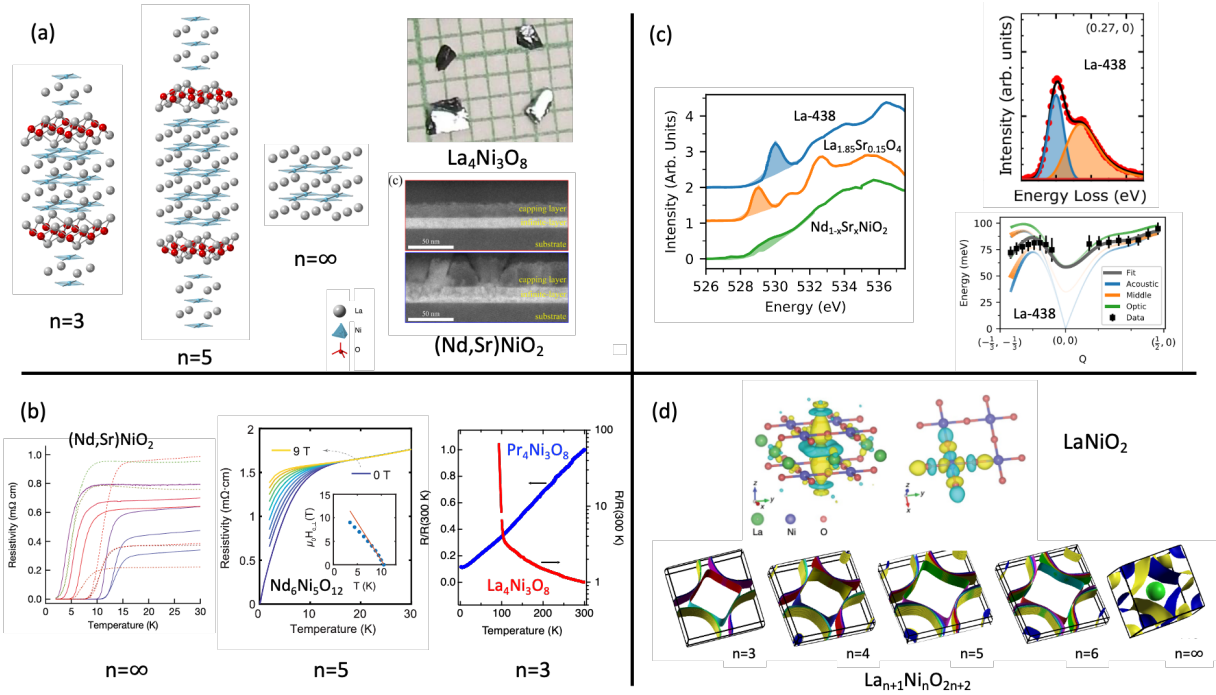

Fig. 1. Overview of $d^9$ nickelates structure and properties (a) left: members of the layered $d^9$ nickelate family showing square planar $NiO_2$ sheets separated by fluorite-like $R_2O_2$ layers; right top: single crystals of $La_4Ni_3O_8$; right bottom: cross-sectional images of PLD-grown (Nd,Sr)NiO$_2$ films (Ref. 46 ). (b) transport properties showing superconductivity in thin film infinite layer (Nd,Sr)NiO$_2$ (Ref. 2) and five-layer $Nd_6Ni_5O_{12}$ (Ref. 3) and competing ground states in single crystal bulk trilayer nickelates, metallic $Pr_4Ni_3O_8$ and charge-stripe insulating $La_4Ni_3O_8$. Note logarithmic scale for La-438. (c) Electronic structure of nickelates. left: prominent O $K$-edge prepeak in cuprates and trilayer $La_4Ni_3O_8$ appears suppressed in (Nd,Sr)NiO$_2$ films (Ref 49). right: RIXS data reveal a magnon whose dispersion can be fit to yield the nearest-neighbor Ni-O superexchange constant (Ref 49). (d) DFT calculations of electronic structure of $d^9$ layered nickelates. top: Wannier orbitals calculated for LaNiO$_2$ showing both $dx^2$-$y^2$ and $dz^2$ configurations (Ref. 51); bottom: calculated Fermi surface vs layer number for the $La_{n+1}Ni_nO_{2n+2}$ series (Ref. 52).

## $d^9$ Nickelates: Materials and Chemistry

If $Ni^{2+}$ isn't right, why not $Ni^{1+}$, which would share the $d^9$ configuration of $Cu^{2+}$? Indeed, Anisimov suggested that $Ni^{1+}$ in the square planar configuration found in the infinite layer compound, if doped with low-spin $Ni^{2+}$, could be a superconductor.[13] In counterpoint, Pickett argued that reduced hybridization and concomitant weighting of the 3$d$ band at $E_F$ meant that despite the isoelectronic configuration, "$Ni^{1+}$ is not $Cu^{2+}$."[14] From the perspective of solid-state inorganic chemistry, Ni prefers to be in the 2+ charge state, and oxides with $Ni^{1+}$ are rare. In the present context, a family of layered $d^9$ compounds has taken center stage (Fig. 1a). This homologous series, $R_{n+1}Ni_nO_{2n+2}$ (R=La,Pr,Nd), first described by Lacorre,[15] is characterized by interleaving blocks of $n$ $NiO_2$ layers separated by $R_2O_2$ fluorite blocks. It is noteworthy that the $n$=1 member of the series is isostructural with the well-known T' phase of the superconducting cuprates, $(Nd,Ce)_2CuO_4$. The remainder of this Perspective focuses on this class of $d^9$ nickelates in both bulk and thin-film forms, their growth, properties, and impact on high $T_c$ superconductivity.

### *Bulk Materials*

The 'infinite layer' nickelate $LaNiO_2$ ($n$=∞, LNO) possesses square planar $NiO_2$ layers, in which the crystal field would produce a hole in the $d_{x^2-y^2}$ band, desirable for cuprate superconductors. Synthesis of LNO in bulk polycrystalline form was reported by Crespin in 1983 (followed up 22 years later) using a complex apparatus to titrate $H_2$ gas to the sample.[16-18] Study of the evolution of $LaNiO_3$ to reduced phases was reported by Moriga, et al.[19-21] who were unable to reach $LaNiO_2$. The synthesis window of temperature, sample size, etc. employed by Crespin was quite narrow, and to our knowledge, no other successful synthesis of LNO in $H_2$ gas has been reported.

Hayward reported bulk synthesis of LNO and $NdNiO_2$ (NNO) using NaH as a reducing agent.[22, 23] The synthesis led to mixed phases with excess O, stacking faults, etc. NNO synthesized in this way was insulating and paramagnetic, not the hoped-for antiferromagnetism found in cuprate parent phases. Hayward also successfully doped 10% Sr and Ca into LNO, but he reported no physical properties due to contamination by metallic Ni byproducts.[24] More recently, Li et al.[25] synthesized bulk $Nd_{1-x}Sr_xNiO_2$ (x = 0, 0.2, 0.4) using $CaH_2$ to reduce the perovskite parent. The samples were insulating and paramagnetic. Contemporaneously, Wang et al.[26] made the x=0.0, 0.1, 0.2 compositions and found no evidence for the mixed phase behavior found by Hayward. Samples were insulating, and neutron diffraction revealed the absence of long-range magnetic order above 3 K in the parent phase. Some theoretical considerations implicate $H^-$ insertion into the infinite layer nickelates during the reduction process.[27, 28] Such interstitial species could dramatically impact the electronic band structure; however, no direct experimental evidence for this interstitial has been reported.

The multilayer analogs of 'infinite layer' $R_{n+1}Ni_nO_{2n+2}$ (R=La,Pr,Nd) shown in Fig. 1a are synthesized by reduction of the corresponding Ruddlesden-Popper phase, $R_{n+1}Ni_nO_{3n+1}$. In bulk materials, samples with La, Pr, and Nd have been synthesized with $n \leq 3$.[29-31] Unlike the infinite layer materials, these compounds can be synthesized at 300 °C using dilute mixtures of $H_2$ in inert gas, or using NaH in organic solvents at a more mild 150 °C, as shown by Poltavets.[30] The ease of reduction can be traced to the mixed-valent nature of this family, which implies a fraction of $Ni^{2+}$ for all $n$. Given the square planar coordination, it was conceivable that the divalent Ni could enter as low spin, satisfying the conditions placed by Anisimov for superconductivity.[13]

Searching for cuprate analogs, particular attention was paid to the $n$=3 material, $La_4Ni_3O_8$ (La-438), which formally contains $Ni^{1+}$ and $Ni^{2+}$ in a 2:1 ratio. Work by Poltavets and Greenblatt[29] revealed a semiconductor to metal transition at T=105 K, which also had a magnetic component. Speculation at the time was that either a CDW or SDW was the cause of this transition, and band structure calculations corroborated this $q$-space mechanism by showing a nested Fermi surface.[29] Other suggestions included charge order among the planes[32]—outer planes $Ni^{1+}$ and inner plane $Ni^{2+}$—or a spin-state transition mediated by inter-plane coupling through partially-occupied $dz^2$ orbitals.[33] It was the successful growth of single crystals at Argonne National Laboratory, using a high oxygen pressure floating zone furnace followed by $H_2$ reduction, that opened a new view on La-438: we showed unambiguously using x-ray and neutron diffraction that the transition results from real space charge- and spin-stripe formation, leading to an insulating ground state (Fig. 1b, right), with a three-cell repeat set by the hole concentration: 1/3 hole doped into a $Ni^{1+}$ background.[34,35] The correspondence to the stripes found in hole-doped $Ni^{2+}$ LSNO (including the diagonal propagation vector) became obvious in hindsight. DFT calculations show that the stripe formation is accompanied by a buckling of the outer Ni-O layers, a prediction verified by high-resolution diffraction measurements.[36] µSR studies by Bernal were found to be consistent with this stripe model.[37]

We were also able to grow the Pr analog, $Pr_4Ni_3O_8$ (Pr-438).[38] As shown in Fig. 1b (right), Pr-438 is metallic. $Pr^{3+}$ is somewhat smaller than $La^{3+}$, and this conceivably leads to a wider bandwidth than that of La-438, favoring the metallic ground state rather than the charge- and spin-stripes found for La-438. Notably, however, RIXS measurements on Pr-438 show the same broadened magnon dispersion as La-438 (see below) despite no long-range magnetic order, implying that stripe correlations are present here as well. As discussed below, Pr-438 possesses many of the ingredients for cuprate superconductivity. However, its hole concentration is slightly beyond the cuprate superconducting dome, and it remains normal. DFT calculations[39] and ARPES measurements[40] substantiate this picture, revealing an electronic structure like that of overdoped cuprates. Unfortunately, attempts to dope electrons into Pr-438 and shift the hole concentration into the superconducting dome have thus far proven unsuccessful.

*Thin Films*

Thin film synthesis of LNO was achieved by Kawai et al.[41,42] using $CaH_2$ reduction of a perovskite precursor film. It was found to be insulating. This result was subsequently confirmed by Kaneko et al.[43] and Ikeda et al.[44,45] Regrettably, these groups do not seem to have pursued hole doping into the parent LNO or the effect of using other rare earths than La on the A site. It thus came as a remarkable surprise in August 2019, when the Stanford team led by Harold Hwang reported superconductivity at ≈15 K in PLD films of $Nd_{1-x}Sr_xNiO_2$ grown on STO substrates and capped by STO (Fig 1b, left).[2] Synthesis of high quality, superconducting films is challenging, requiring optimization of multiple growth parameters to avoid competing phases of multilayer nickelates and other second phases (Fig 1a, right bottom).[46] While the sample-to-sample variation of $T_c$ shown in these early samples reflects subtle compositional or structural variables, subsequent work has refined the synthesis procedure and revealed a $T_c(x)$ that shows a superconducting dome reminiscent of, and at similar hole concentrations as, the cuprates. Subsequently, other groups have confirmed the Stanford result, and have extended it to show superconductivity in R=La, and Pr analogs.[47] With the Stanford group's announcement, the 30+ year pursuit of nickelate superconductivity had finally reached its hoped-for end, but a new quest for understanding this novel superconductor and its relationship to copper oxides had just begun.

Following on the direction suggested by the bulk trilayer nickelates, a strategy of growing multilayer thin-film nickelates has been pursued by Julia Mundy's group at Harvard through another tour de force in materials growth.[3] While in infinite layer (Nd,Sr)NiO$_2$ and the putative trilayer $(Pr,Ce)_4Ni_3O_8$ the doping is provided by aliovalent substitution on the A site, it is also possible to self-dope the higher order members of the $R_{n+1}Ni_nO_{2n+2}$ series, as the formal Ni oxidation state follows $(n+1)/n$. Thus, by stacking more layers, the average Ni oxidation state will progressively move into the superconducting dome of the cuprates and the infinite-layer nickelates, entering perhaps at $n=4$ and certainly by $n=5$ (See Fig. 2). This hypothesis proved to be correct, as shown in Fig. 1b (middle). Here the resistivity of the five-layer material, $Nd_6Ni_5O_{12}$ (nominal Ni oxidation state = 6/5), is shown as a function of temperature and field. A broad superconducting transition is found with $T_c$ (H=0) around 10 K, making this multilayer, self-doped material the second class of $d^9$-derived nickelate superconductors and validating a unified a picture of nickelate and cuprate superconducting phase diagrams controlled by $d$ electron count.

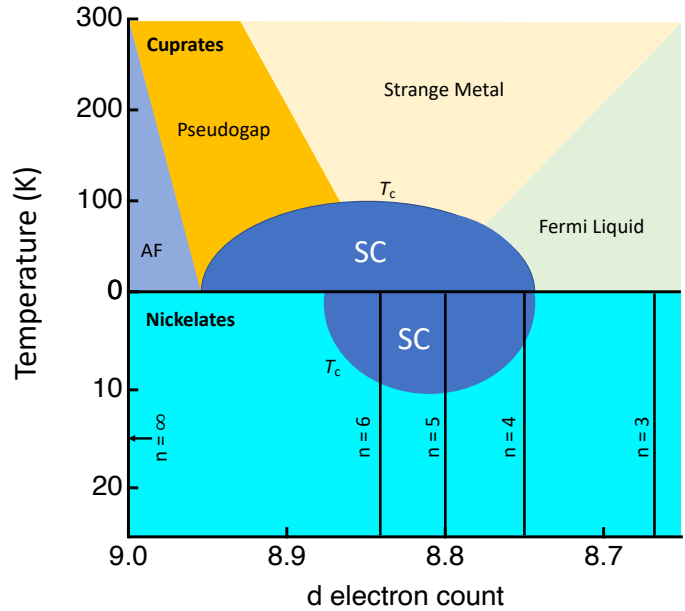

Figure 2 Schematic unified electronic phase diagram of $d^9$ nickelates and cuprates as a function of $d$ electron count. Ispired by Refs. 3, 4 and 48.

## $d^9$ Nickelates: Physics

We now turn from materials and chemistry to physics to overview the current thinking about the question of how nickelates compare and contrast with the cuprates. Inspired by the cuprate ingredients discussed above, we touch on magnetism and electronic structure and the role of hybridization on the hole wavefunction.

### *Magnetism*

The cuprate parent phase is an antiferromagnetic insulator. As mentioned above, this does not seem to be the case for the infinite layer nickelates, which in bulk are paramagnetic.[23, 25, 26] Nonetheless, Ni $L_3$-edge RIXS measurements on the superconducting films by Lu et al.[48] find a magnon spectrum consistent with the spin-1/2 antiferromagnetic square lattice found in cuprates. The nearest-neighbor exchange, $J_1$=63 meV, is roughly half that found in copper oxide materials. Mark Dean's group at Brookhaven National Lab has analyzed similar experiments on single crystals of trilayer La-438 (Fig 1c, right), finding a comparable value of $J_1$=69 meV, although the nominal doping concentration is quite different.[49] A broadened magnon spectrum observed by this group is consistent with the spin stripes seen by neutrons. It is currently impossible to synthesize the equivalent hole concentration in the trilayer material in bulk or to reach the parent (e.g., $Pr_3CeNi_3O_8$), but DFT calculations indicate the antiferromagnetic square lattice as the ground state of this putative parent phase.[39]

*Electronic Structure*

DFT calculations show that the broad features of the infinite layer nickelate Fermi surface to be much like that of the copper oxides, except for the rare earth $d$-band hole pocket that self-dopes the former.[14, 50] The size of this hole pocket decreases with doping into the superconducting dome. The Wannier orbitals shown in Fig. 1d (top) bear strong resemblance to the one-electron $dx^2-y^2$ and $dz^2$ orbitals expected from simple crystal field considerations.[51] Looking at the multilayer systems (Fig. 1 (bottom)), the number of bands in the $dx^2-y^2$ manifold increases with $n$, as expected.[52]

Whether the nickelates lie in the Mott-Hubbard or charge transfer regime remains a significant controversy in the field. Since before even the earliest mentions of superconductivity in infinite layer nickelates, a Mott picture has been favored due to the larger $3d$-O $2p$ separation.[14] Modern DFT calculations on the $d^9$ nickelates support this picture,[53-56] which leads to less metal-ligand hybridization, and to the inference that Mott physics should be a better descriptor than a charge-transfer insulator model, distancing the $d^9$ nickelates from cuprates. Experimental support for this picture comes from STEM-EELS measurements that find $NdNiO_2$ lies in the Mott regime and that with doping, holes appear on the Ni and Nd $d$ bands as well as on O.[53] Many-body DMFT calculations have come down on both sides of this divide, but recent calculations seem to favor the Mott picture rather than charge-transfer, which is found for the isostructural cuprate analog $CaCuO_2$.[56] Experimentally, the prominent O $2p$ $K$-edge prepeak found in cuprates that signifies strong $3d$-O $2p$ hybridization is greatly suppressed in the infinite layer materials, although less so in the case of the trilayer material (Fig. 1c, left). Adding to the uncertainty, the Brookhaven group argues from O $K$-edge RIXS and cluster calculations that neither limit is appropriate for the trilayer La-438; rather, this nickelate is better described as having a mixed Mott-Hubbard/charge-transfer character with significant hole character on O, explaining the large superexchange mentioned above in this material.[57] The jury is still out on hybridization in the $d^9$ nickelates, but it seems that oxygen states cannot be ignored in a proper description of these materials.

To date, the gold-standard for probing the low-energy physics of copper oxide superconductors, Angle-Resolved Photoemission Spectroscopy (ARPES), has not been applied to the thin film superconducting nickelates. It has, however, been measured on single crystals of the trilayer metal Pr-438 by the Dessau group. In unpublished data, they find a Fermi surface consistent with that of overdoped BSCCO and a large mass enhancement signaling strong correlations.[40]

Finally, the crystal field is expected to strongly favor the low-spin configuration on $Ni^{2+}$. This configuration has indeed been confirmed in the trilayer materials by x-ray absorption spectroscopy, which also shows a strong in plane orbital polarization.[38] Both findings stand as analogous to the cuprate electronic state.

**Outlook**

To conclude this Perspective, a discussion of open questions and potential directions to pursue is appropriate. Here are a few materials-inspired questions that could direct research in the near term.

*Why aren't bulk $d^9$ nickelates superconducting?*

To date, no superconductivity has been found in bulk infinite layer phases at doping levels inside the superconducting dome of thin film specimens or in bulk multilayer systems ($n$=2,3) that lie outside the dome. This does not mean it is impossible, so what might be the problem? It could be as simple as having the wrong lattice constant. The films are slightly compressively strained vs. freestanding bulk materials, and so perhaps pressure could tip the balance. Unfortunately, pressure studies on $Nd_{0.8}Sr_{0.2}NiO_2$ to 50 GPa show no sign of metallic, let alone superconducting, behavior.[25] Concerning is the structural perfection[2, 46, 53] found in the ultra-thin films that is probably unlikely to be found in bulk materials. Oxygen defects may become disordered or be insufficiently ordered during the low-T reduction. The complex process that takes $RNiO_3$ to $RNiO_2$ passes through multiple known (and perhaps unknown) ordered O vacancy structures.[19-21, 58] And while single crystals have been made by $CaH_2$ reduction of $Nd_{0.8}Ca_{0.2}NiO_3$ grown at high P, and they are more metallic than powder samples, there is a low-temperature localization transition and no superconductivity.[59] This may reflect a true Ca content in the crystals that is insufficient to reach the superconducting dome. Further work here is called for.

The message is that we simply may not have optimized materials growth of bulk samples to the same level of perfection as the films. Desperately needed is a direct synthesis of the $d^9$ nickelate families that does not involve the perovskite reduction process. Such a method per force must be low temperature, as the infinite layer materials begin to disorder and to decompose at or below ~300 °C, depending on the rare earth. Such a creative breakthrough in synthesis and growth would be game-changing.

*Can trilayer materials be made SC?*

Based on simple $d$ count as the key parameter, the trilayers materials are overdoped by a bit (Fig. 2). Substituting on the Pr site of Pr-438 with a tetravalent ion like $Ce^{4+}$ could in principle adjust the hole count into the dome. Despite extensive efforts, we have failed in this endeavor (A small amount of $Ce^{4+}$ could be substituted into La-438, causing the metal-insulator transition to shift down from 105 K by 25 K). Other methods could include electric field gating or K-dosing (e.g., in an ARPES chamber) as done for cuprates and iridates.[60, 61] So far no one has succeeded at these strategies. A contrarian approach has been suggested by Dessau: while the focus has been on trying to increase the electron count to bring Pr-438 into the dome, another possibility is that the $dz^2$ band takes up some holes that could need to be in the $dx^2-y^2$ band, and that hole doping is really needed to tune into superconductivity. However, the recent report of superconductivity in $n$=5 $Nd_6Ni_5O_{12}$ films by Mundy et al.[3] (Fig. 1b and Fig. 2) argues that the electron doping argument is appropriate, although this does not rule out other approaches. Even in the absence of superconductivity, the trilayer materials may hold other interesting stories. For instance, how does the stripe phase of La-438 evolve into the correlated metal Pr-438 at T=0 K?

*What other prospects may be out there?*

We have argued in this Perspective that materials synthesis and crystal growth advances have led the way in the rebirth of nickelates. Largely this has emerged as we access and gain better control over materials with Ni in extreme oxidation states spanning from +1 to +4. These synthesis advances have largely been applied to known materials. We should also look to the possibility of other materials — either known or unknown — in the quest for a more complete picture of nickelate superconductivity and the more general understanding of the physics of these highly correlated materials. Hybrid materials such as layered oxide-chalcogenides stand out as an interesting direction to pursue.

As demonstrated by the intense study of LSNO and related $Ni^{2+}$ materials and the long history of the $RNiO_3$ $Ni^{3+}$ perovskites, the rare earth nickelates as an extended family are more than just the $d^9$ $Ni^{1+}$ materials, and many open questions remain. What is the magnetic ground state of the rare earth perovskites? Do dynamic fluctuations exist in the metallic phase of $LaNiO_3$? How does dimensionality evolution from LSNO to $LaNiO_3$ influence the transition from a real space charge stripe insulator to a metal? Many if not all such questions stand at a better place to be answered than they did even a few years ago because of renewed interest and capability in growing crystals under high pressure either by zone techniques or high fugacity solution growth in anvil cells or optimizing thin film growth. We are already seeing new and yet unresolved physics in these crystals: a broad maximum in magnetic susceptibility unknown in powders is universally found in $LaNiO_3$ crystals grown independently by multiple groups around the world.[62-64] Evidence for a polar distortion in perovskite nickelates is growing.[65] A coupled CDW/SDW in the trilayer

Ruddlesden-Popper phase $La_4Ni_3O_{10}$ replaces the charge stripes in single layer $La_{2-x}Sr_xNiO_4$, hinting at an evolution between $q$-space and real space driven instabilities tuned by dimensionality.[66]

The nickelate field had already established its longevity and importance to condensed matter science well before superconductivity was discovered in 2019, and the next few years should prove exciting as we learn more about superconducting and non-superconducting systems alike. As we master the synthesis and growth of a wider range of nickel oxides in extreme oxidation states, it is certain that new answers to questions both old and new will be found and that unexpected surprises are virtually guaranteed to emerge.

There is much work to be done. So, let the fields lay fallow, but not for too long.

Acknowledgement: This work was sponsored by the U.S. Department of Energy, Office of Science, Basic Energy Sciences, Materials Science and Engineering Division. The author thanks Mike Norman and Daniel Phelan for helpful discussions in the preparation of this manuscript.

References:

1. J. Dewey, *Art as Experience*. (Minton, Balch & Company, New York, 1934).

2. D. Li, K. Lee, B. Y. Wang, M. Osada, S. Crossley, H. R. Lee, Y. Cui, Y. Hikita and H. Y. Hwang, Nature **572**, 624 (2019).

3. G. A. Pan, D. F. Segedin, H. LaBollita, Q. Song, E. M. Nica, B. H. Goodge, A. T. Pierce, S. Doyle, S. Novakov, D. C. Carrizales, A. T. N'Diaye, P. Shafer, H. Paik, J. T. Heron, J. A. Mason, A. Yacoby, L. F. Kourkoutis, O. Erten, C. M. Brooks, A. S. Botana and J. A. Mundy, arXiv:2109.09726 (2021).

4. B. Keimer, S. A. Kivelson, S. A. Kivelson, M. R. Norman, S. Uchida, S. Uchida, J. Zaanen and J. Zaanen, Nature **518**, 179 (2015).

5. J. Zaanen, G. A. Sawatzky and J. W. Allen, Phys. Rev. Lett. **55**, 418 (1985).

6. S. Shinomori, Y. Okimoto, M. Kawasaki and Y. Tokura, J. Phys. Soc. Jpn. **71**, 705 (2002).

7. S. W. Cheong, H. Y. Hwang, C. H. Chen, B. Batlogg, L. W. Rupp and S. A. Carter, Phys. Rev. B **49**, 7088 (1994).

8. T. Katsufuji, T. Tanabe, T. Ishikawa, S. Yamanouchi, Y. Tokura, T. Kakeshita, R. Kajimoto and H. Yoshizawa, Phys. Rev. B **60**, R5097 (1999).

9. J. M. Tranquada, D. J. Buttrey and V. Sachan, Phys. Rev. B **54**, 12318 (1996).

10. P. Wochner, J. M. Tranquada, D. J. Buttrey and V. Sachan, Phys. Rev. B **57**, 1066 (1998).

11. A. T. Boothroyd, D. Prabhakaran, P. G. Freeman, S. J. S. Lister, M. Enderle, A. Hiess and J. Kulda, Phys. Rev. B **67**, 100407 (2003).

12. H. Yoshizawa, T. Kakeshita, R. Kajimoto, T. Tanabe, T. Katsufuji and Y. Tokura, Phys. Rev. B **61**, R854 (2000).

13. V. I. Anisimov, D. Bukhvalov and T. M. Rice, Phys. Rev. B **59**, 7901 (1999).

14. K. W. Lee and W. E. Pickett, Phys. Rev. B **70**, 165109 (2004).

15. P. Lacorre, J. Solid State Chem. **97**, 495 (1992).

16. M. Crespin, O. Isnard, F. Dubois, J. Choisnet and P. Odier, J. Solid State Chem. **178**, 1326 (2005).

17. M. Crespin, P. Levitz and L. Gatineau, J. Chem. Soc. Faraday Trans. 2 **79**, 1181 (1983).

18. P. Levitz, M. Crespin and L. Gatineau, J. Chem. Soc. Faraday Trans. 2 **79**, 1195 (1983).

19. T. Moriga, O. Usaka, I. Nakabayashi, T. Kinouchi, S. Kikkawa and F. Kanamaru, Solid State Ionics **79**, 252 (1995).

20. T. Moriga, O. Usaka, I. Nakabayashi, Y. Hirashima, T. Kohno, S. Kikkawa and F. Kanamaru, Solid State Ionics **74**, 211 (1994).

21. T. Moriga, O. Usaka, T. Imamura, I. Nakabayashi, I. Matsubara, T. Kinouchi, S. Kikkawa and F. Kanamaru, Bull. Chem. Soc. Jpn. **67**, 687 (1994).

22. M. A. Hayward, M. A. Green, M. J. Rosseinsky and J. Sloan, J. Amer. Chem. Soc. **121**, 8843 (1999).

23. M. A. Hayward and M. J. Rosseinsky, Solid State Sciences **5**, 839 (2003).

24. M. A. Hayward, University of Oxford (1999).

25. Q. Li, C. He, J. Si, X. Zhu, Y. Zhang and H.-H. Wen, Commun. Mat. **1**, 16 (2020).

26. B.-X. Wang, H. Zheng, E. Krivyakina, O. Chmaissem, P. P. Lopes, J. W. Lynn, L. C. Gallington, Y. Ren, S. Rosenkranz, J. F. Mitchell and D. Phelan, Phys. Rev. Mat. **4**, 084409 (2020).

27. L. Si, W. Xiao, J. Kaufmann, J. M. Tomczak, Y. Lu, Z. Zhong and K. Held, Phys. Rev. Lett. 124, 166402 (2020).

28. O. I. Malyi, J. Varignon and A. Zunger, arXiv:2107.01790 (2021).

29. V. V. Poltavets, K. A. Lokshin, A. H. Nevidomskyy, M. Croft, J. Hadermann, G. V. Tendeloo, T. Egami and M. Greenblatt, Phys. Rev. Lett. **104**, 206403 (2010).

30. C. K. Blakely, S. R. Bruno and V. V. Poltavets, Inorg. Chem. **50**, 6696 6700 (2011).

31. Y. Sakurai, N. Chiba, Y. Kimishima and M. Uehara, Physica C **487**, 27(2013).

32. H. Wu, New J. Phys. **15**, 023038 (2013).

33. V. Pardo and W. E. Pickett, Phys. Rev. Lett. **105**, 266402 (2010).

34. J. Zhang, Y.-S. Chen, D. Phelan, H. Zheng, M. R. Norman and J. F. Mitchell, Proc. Natl. Acad. Sci. USA **113**, (2016).

35. J. Zhang, D. M. Pajerowski, A. S. Botana, H. Zheng, L. Harriger, J. Rodriguez-Rivera, J. P. C. Ruff, N. J. Schreiber, B. Wang, Y.-S. Chen, W. C. Chen, M. R. Norman, S. Rosenkranz, J. F. Mitchell and D. Phelan, Phys. Rev. Lett. **122**, 247201 (2019).

36. A. S. Botana, V. Pardo, W. E. Pickett and M. R. Norman, Phys. Rev. B **94**, 081105 (2016).

37. O. O. Bernal, D. E. MacLaughlin, G. D. Morris, P. C. Ho, L. Shu, C. Tan, J. Zhang, Z. Ding, K. Huang and V. V. Poltavets, Phys. Rev. B **100**, 125142 (2019).

38. J. Zhang, A. S. Botana, J. W. Freeland, D. Phelan, H. Zheng, V. Pardo, M. R. Norman and J. F. Mitchell, Nature Physics **13**, 864 (2017).

39. A. S. Botana, V. Pardo and M. R. Norman, Phys. Rev. Materials **1**, 021801 (2017).

40. H. Li, P. Hao, J. Zhang, K. Gordon, A. G. Linn, H. Zheng, X. Zhou, J. F. Mitchell and D. S. Dessau, submitted.

41. M. Kawai, S. Inoue, M. Mizumaki, N. Kawamura, N. Ichikawa and Y. Shimakawa, Applied Physics Lett. **94**, 082102 (2009).

42. M. Kawai, K. Matsumoto, N. Ichikawa, M. Mizumaki, O. Sakata, N. Kawamura, S. Kimura and Y. Shimakawa, Cryst. Growth Des. **10**, 2044 (2010).

43. D. Kaneko, K. Yamagishi, A. Tsukada, T. Manabe and M. Naito, Physica C: Superconductivity **469**, 936 (2009).

44. A. Ikeda, Y. Krockenberger, H. Irie, M. Naito and H. Yamamoto, Appl. Phys. Express **9**, 061101 (2016).

45. A. Ikeda, T. Manabe and M. Naito, Physica C: Superconductivity and its Applications **506**, 83 (2014).

46. K. Lee, B. H. Goodge, D. Li, M. Osada, B. Y. Wang, Y. Cui, L. F. Kourkoutis and H. Y. Hwang, APL Mater. **8**, 041107 (2020).

47. H. Lin, D. J. Gawryluk, Y. M. Klein, S. Huangfu, E. Pomjakushina, F. v. Rohr and A. Schilling, arXiv:2104.14324 (2021).

48. H. Lu, M. Rossi, A. Nag, M. Osada, D. F. Li, K. Lee, B. Y. Wang, M. Garcia-Fernandez, S. Agrestini, Z. X. Shen, E. M. Been, B. Moritz, T. P. Devereaux, J. Zaanen, H. Y. Hwang, K.-J. Zhou and W. S. Lee, Science **373**, 213 (2021).

49. J. Q. Lin, P. V. Arribi, G. Fabbris, A. S. Botana, D. Meyers, H. Miao, Y. Shen, D. G. Mazzone, J. Feng, S. G. Chiuzbaian, A. Nag, A. C. Walters, M. García-Fernández, K. J. Zhou, J. Pelliciari, I. Jarrige, J. W. Freeland, J. Zhang, J. F. Mitchell, V. Bisogni, X. Liu, M. R. Norman and M. P. M. Dean, Phys. Rev. Lett. **126**, 087001 (2021).

50. M.-Y. Choi, K.-W. Lee and W. E. Pickett, Phys. Rev. B **101**, 020503 (2020).

51. M. Hepting, D. Li, C. J. Jia, H. Lu, E. Paris, Y. Tseng, X. Feng, M. Osada, E. Been, Y. Hikita, Y. D. Chuang, Z. Hussain, K. J. Zhou, A. Nag, M. Garcia-Fernandez, M. Rossi, H. Y. Huang, D. J. Huang, Z. X. Shen, T. Schmitt, H. Y. Hwang, B. Moritz, J. Zaanen, T. P. Devereaux and W. S. Lee, Nature Materials, **19**, 381 (2020).

52. H. LaBollita and A. S. Botana, Phys. Rev. B **104**, 035148 (2021).

53. B. H. Goodge, D. Li, K. Lee, M. Osada, B. Y. Wang, G. A. Sawatzky, H. Y. Hwang and L. F. Kourkoutis, Proceedings of the National Academy of Sciences **118**, e2007683118 (2021).

54. M. Hepting, R. J. Green, Z. Zhong, M. Bluschke, Y. E. Suyolcu, S. Macke, A. Frano, S. Catalano, M. Gibert, R. Sutarto, F. He, G. Cristani, G. Logvenov, Y. Wang, P. A. v. Aken, P. Hansmann, M. L. Tacon, J.-M. Triscone, G. A. Sawatzky, B. Keimer and E. Benckiser, Nature Physics **14**, 1097 (2019).

55. M. Jiang, M. Berciu and G. A. Sawatzky, Phys. Rev. Lett. **124**, 207004 (2020).

56. J. Karp, A. Hampel, M. Zingl, A. S. Botana, H. Park, M. R. Norman and A. J. Millis, Phys. Rev. B **102**, 245130 (2020).

57. Y. Shen, J. Sears, G. Fabbris, J. Li, J. Pelliciari, I. Jarrige, X. He, I. Bozovic, M. Mitrano, J. Zhang, J. F. Mitchell, A. S. Botana, V. Bisogni, M. R. Norman, S. Johnston and M. P. M. Dean, arXiv:2110.08937 (2021)

58. B.-X. Wang, S. Rosenkranz, X. Rui, J. Zhang, F. Ye, H. Zheng, R. F. Klie, J. F. Mitchell and D. Phelan, Phys. Rev. Mat. **2**, 064404 (2018).

59. P. Puphal, Y.-M. Wu, K. Fürsich, H. Lee, M. Pakdaman, J. A. N. Bruin, J. Nuss, Y. E. Suyolcu, P. A. v. Aken, B. Keimer, M. Isobe and M. Hepting, arXiv:2106.13171 (2021).

60. M. A. Hossain, J. D. F. Mottershead, D. Fournier, A. Bostwick, J. L. McChesney, E. Rotenberg, R. Liang, W. N. Hardy, G. A. Sawatzky, I. S. Elfimov, D. A. Bonn and A. Damascelli, Nature Physics **4**, 527 (2008).

61. Y. K. Kim, N. H. Sung, J. D. Denlinger and B. J. Kim, Nature Physics **12**, 37 (2016).

62. K. Dey, W. Hergett, P. Telang, M. M. Abdel-Hafiez and R. Klingeler, J Cryst Growth **524**, 125157 (2019).

63. H. Guo, Z. W. Li, L. Zhao, Z. Hu, C. F. Chang, C. Y. Kuo, W. Schmidt, A. Piovano, T. W. Pi, O. Sobolev, D. I. Khomskii, L. H. Tjeng and A. C. Komarek, Nature Commun. **9**, 43 (2017).

64. J. Zhang, H. Zheng, Y. Ren and J. F. Mitchell, Cryst. Growth. Des. **17**, 2730 (2017).

65. I. Ardizzone, J. Teyssier, I. Crassee, A. B. Kuzmenko, D. G. Mazzone, D. J. Gawryluk, M. Medarde and D. v. d. Marel, Phys. Rev. Res. **3**, 033007 (2021).

66. J. Zhang, D. Phelan, A. S. Botana, Y.-S. Chen, H. Zheng, M. Krogstad, S. G. Wang, Y. Qiu, J. A. Rodriguez-Rivera, R. Osborn, S. Rosenkranz, M. R. Norman and J. F. Mitchell, Nature Commun. **11**, 6003 (2020).